\documentclass[english,aps,preprint]{revtex4}
\usepackage{pslatex}
\usepackage[T1]{fontenc}
\usepackage[latin1]{inputenc}

\makeatletter

\usepackage{babel}
\makeatother
\begin{document}

\preprint{Aug 31, 2005}

\title{The Lattice Green Function for the Poisson Equation on an Infinite
Square Lattice}

\author{Stefan Hollos}

\email{stefan@exstrom.com}

\homepage{http://www.exstrom.com/stefan/stefan.html}

\affiliation{Exstrom Laboratories LLC, 662 Nelson Park Dr, Longmont, Colorado
80503, USA.}

\author{Richard Hollos}

\affiliation{Exstrom Laboratories LLC, 662 Nelson Park Dr, Longmont, Colorado
80503, USA.}

\begin{abstract}
We derive formulas for the matrix elements of the lattice Green function
for the discrete Poisson equation on an infinite square lattice. The
partial difference equation for the matrix elements is solved by reducing
it to a series of first order difference equations, which can then
be solved sequentially. These formulas are useful in solving two dimensional
Poisson equation problems using the finite difference approximation.
\end{abstract}
\maketitle

\section{introduction}

In this paper we will derive formulas for the matrix elements of the
lattice Green function for the discrete Poisson equation on an infinite
square lattice. The discrete Poisson equation arises when the finite
difference approximation is applied to the continuous Poisson equation.
The results derived here will therefore be of considerable interest
for solving any problem that can be modeled by a two dimensional Poisson
equation \cite{cserti00}. One particularly large area of application
is in the solution of two dimensional electrostatics and magnetostatics
problems. Calculating capacitance, inductance, and charge distributions
on conductors with boundary conditions at infinity are problems of
great practical as well as some theoretical interest \cite{exstrom2005}. 

We will begin by introducing some notation and a formulation of the
problem. We let $L$ denote the lattice Laplacian operator and then
write the discrete Poisson equation as:\begin{equation}
L\vert\phi\rangle=\vert f\rangle\label{eq:1}\end{equation}
In the lattice basis we have basis vectors $\vert n\rangle$ associated
with the lattice point $\vec{r}_{n}=n_{1}\vec{a}_{1}+n_{2}\vec{a}_{2}$,
where $\vec{a}_{i}=a\hat{x}_{i}$, $\hat{x}_{i}\cdot\hat{x}_{j}=\delta(i,j)$,
and $n_{i}$ is an integer. In this basis eq. \ref{eq:1} becomes\begin{equation}
\sum_{n}L_{ln}\phi(\vec{r}_{n})=f(\vec{r}_{l})\label{eq:2}\end{equation}
 The Laplacian matrix elements are given by\begin{equation}
L_{ln}=-4\delta(\vec{r_{l},}\vec{r}_{n})+\delta(\vec{r}_{l}+\vec{a}_{1},\vec{r}_{n})+\delta(\vec{r}_{l}-\vec{a}_{1},\vec{r}_{n})+\delta(\vec{r}_{l}+\vec{a}_{2},\vec{r}_{n})+\delta(\vec{r}_{l}-\vec{a}_{2},\vec{r}_{n})\label{eq:3}\end{equation}
The Green function is defined by the equation $LG=-I$ which in the
lattice basis is\begin{equation}
\sum_{n}L_{ln}G_{nm}=\sum_{n}L(\vec{r}_{l}-\vec{r}_{n})G(\vec{r}_{n}-\vec{r}_{m})=-\delta(l,m)\label{eq:4}\end{equation}
Using eq. \ref{eq:3} gives the following recurrence for the matrix
elements of $G$.\begin{equation}
4G(\vec{r}_{l}-\vec{r}_{m})-\sum_{i=1}^{2}\left[G(\vec{r}_{l}+\vec{a}_{i}-\vec{r}_{m})+G(\vec{r}_{l}-\vec{a}_{i}-\vec{r}_{m})\right]=\delta(l,m)\label{eq:5}\end{equation}
With the notation $\vec{r}_{l}-\vec{r}_{m}=(l_{1}-m_{1})\vec{a}_{1}+(l_{2}-m_{2})\vec{a}_{2}=p_{1}\vec{a}_{1}+p_{2}\vec{a}_{2}$,
and $G(\vec{r}_{l}-\vec{r}_{m})=G(p_{1},p_{2})$, eq. \ref{eq:5}
becomes\begin{equation}
4G(p_{1},p_{2})-G(p_{1}+1,p_{2})-G(p_{1}-1,p_{2})-G(p_{1},p_{2}+1)-G(p_{1},p_{2}-1)=\delta(p_{1},0)\delta(p_{2},0)\label{eq:6}\end{equation}
For an infinite lattice, an eigenbasis expansion gives the following
formula for the matrix elements of $G$ \cite{hollos05_2}\begin{equation}
G(p_{1},p_{2})=\frac{1}{2\pi^{2}}\int_{0}^{\pi}\int_{0}^{\pi}\frac{\cos x_{1}p_{1}\cos x_{2}p_{2}}{2-\cos x_{1}-\cos x_{2}}dx_{1}dx_{2}\label{eq:7}\end{equation}
The problem with this equation is that the integral is divergent for
all values of $p_{1}$and $p_{2}$. We get around this problem by
using the origin referenced Green function $g(p_{1},p_{2})=G(0,0)-G(p_{1},p_{2})$
which is then given by the following integral\begin{equation}
g(p_{1},p_{2})=\frac{1}{2\pi^{2}}\int_{0}^{\pi}\int_{0}^{\pi}\frac{1-\cos x_{1}p_{1}\cos x_{2}p_{2}}{2-\cos x_{1}-\cos x_{2}}dx_{1}dx_{2}\label{eq:8}\end{equation}
This integral is finite for all finite values of $p_{1}$ and $p_{2}$.

$g(p_{1},p_{2})$ will also provide a solution to eq. \ref{eq:1}
provided that the sum of the source terms, $f(\vec{r}_{n})$ over
all the lattice sites is equal to zero. To see this, note that the
solution to eq. \ref{eq:1} in the lattice basis is given by\begin{equation}
\phi(\vec{r}_{l})=-\sum_{n}G_{ln}f(\vec{r}_{n})\label{eq:9}\end{equation}
Now if we have\begin{equation}
\sum_{n}f(\vec{r}_{n})=0\label{eq:10}\end{equation}
Then clearly\begin{equation}
\phi(\vec{r}_{l})=\sum_{n}g_{ln}f(\vec{r}_{n})\label{eq:11}\end{equation}

Since the coefficients of the recurrence in eq. \ref{eq:6} add to
zero we see that $g(p_{1},p_{2})$ must obey the equation\begin{equation}
4g(p_{1},p_{2})-g(p_{1}+1,p_{2})-g(p_{1}-1,p_{2})-g(p_{1},p_{2}+1)-g(p_{1},p_{2}-1)=-\delta(p_{1},0)\delta(p_{2},0)\label{eq:12}\end{equation}
We will now show how to solve this equation for $g(p_{1},p_{2})$.

\section{general solution of the difference equation}

The key to solving eq. \ref{eq:12} is the use of the symmetry $g(p_{1},p_{2})=g(p_{2},p_{1})$.
This will allow eq. \ref{eq:12} to be reduced to a series of first
order difference equations for the matrix elements along the subdiagonals.
The value of the elements along the main diagonal will then provide
a unique solution to the problem. These values can be directly calculated
via eq. \ref{eq:8} \cite{hollos05_1}.

We begin by letting $p_{1}=p_{2}=n$ in eq. \ref{eq:12} and using
the fact that $g(p_{1},p_{2})=g(p_{2},p_{1})$, we get\begin{equation}
g(n+1,n)+g(n,n-1)=2g(n,n)+\frac{1}{2}\delta(n,0)\label{eq:13}\end{equation}
This equation is really just a first order difference equation for
the first subdiagonal matrix elements. To make this observation explicit
we introduce the following notation: $r_{0}(n)=g(n,n)$, and $r_{1}(n)=g(n,n-1)$,
$r_{1}(n+1)=g(n+1,n)$. Eq. \ref{eq:13} then becomes\begin{equation}
r_{1}(n+1)+r_{1}(n)=2r_{0}(n)+\frac{1}{2}\delta(n,0)\label{eq:14}\end{equation}
This equation for the first subdiagonal matrix elements, $r_{1}(n)$,
is easily solved once the diagonal matrix elements, $r_{0}(n)$, are
known. The initial condition for the equation is $r_{1}(1)=r_{1}(0)$.
This comes from the definition, $r_{1}(n)=g(n,n-1)$, from which we
get $r_{1}(1)=g(1,0)$ and $r_{1}(0)=g(0,-1)=g(1,0)$. The value of
the diagonal matrix elements was previously determined to be \cite{hollos05_1}\begin{equation}
g(n,n)=\frac{1}{\pi}\sum_{k=1}^{n}\frac{1}{2k-1}\label{eq:15}\end{equation}
 The solution to eq. \ref{eq:14} can of course be found as the sum
of the solutions to the two equations:\begin{equation}
r_{1}(n+1)+r_{1}(n)=2r_{0}(n)\label{eq:16A}\end{equation}
\begin{equation}
r_{1}(n+1)+r_{1}(n)=\frac{1}{2}\delta(n,0)\label{eq:16B}\end{equation}
The solution that comes from eq. \ref{eq:16A} is the complementary
solution of eq. \ref{eq:12}. The solution corresponding to eq. \ref{eq:16B}
will have $r_{0}(n)=0$ for all $n$.

For the next subdiagonal we let $p_{1}=n$ and $p_{2}=n-1$ in eq.
\ref{eq:12} and we let $r_{2}(n)=g(n,n-2)$ to get:\begin{equation}
r_{2}(n+1)+r_{2}(n)=4r_{1}(n)-r_{0}(n)-r_{0}(n-1)\label{eq:17}\end{equation}
and in general for the $k$th subdiagonal elements we will have \begin{equation}
r_{k}(n+1)+r_{k}(n)=4r_{k-1}(n)-r_{k-2}(n)-r_{k-2}(n-1)\label{eq:18}\end{equation}
where $r_{k}(n)=g(n,n-k)$. This equation can easily be solved to
give $r_{k}(n)$ in terms of $r_{k-1}(n)$ and $r_{k-2}(n)$. 

\begin{equation}
r_{k}(n)=(-1)^{n-k+1}(r_{k-2}(k-2)+r_{k-2}(k-1))-r_{k-2}(n-1)-4\sum_{j=1}^{n-1}(-1)^{j}r_{k-1}(n-j)\label{eq:19}\end{equation}
We apply this equation iteratively with $r_{1}(n)$ given by either
eq. \ref{eq:16A} or eq. \ref{eq:16B} to find a closed form solution
for $r_{k}(n)$. First we look at the case where $r_{1}(n)$ is given
by eq. \ref{eq:16B}. This leads to the following solution:\begin{eqnarray}
r_{0}(n) & = & 0\label{eq:20}\\
r_{1}(n) & = & (-1)^{n+1}\frac{1}{4}\nonumber \\
r_{2}(n) & = & (-1)^{n}(n-1)\nonumber \\
 & \vdots\nonumber \\
r_{k}(n) & = & (-1)^{n+k}\frac{1}{4}\sum_{j=0}^{\lfloor\frac{k-1}{2}\rfloor}4^{k-2j-1}\begin{array}{c}
\left(\begin{array}{c}
k-j-1\\
j\end{array}\right)\left(\begin{array}{c}
n-j-1\\
k-2j-1\end{array}\right)\end{array}\nonumber \end{eqnarray}
At this point we introduce a new function $s(n,m)$, defined as\begin{equation}
s(n,m)=\sum_{j=0}^{\lfloor\frac{n}{2}\rfloor}4^{n-2j}\begin{array}{c}
\left(\begin{array}{c}
n-j\\
j\end{array}\right)\left(\begin{array}{c}
n-j+m/2\\
j+m/2\end{array}\right)\end{array}\label{eq:21}\end{equation}
In terms of this new function, eq. \ref{eq:20} becomes\begin{equation}
r_{k}(n)=(-1)^{n+k}\frac{1}{4}s(k-1,2(n-k))\label{eq:22}\end{equation}
Now the complementary solution, that starts from eq. \ref{eq:16A},
can also be expressed in terms of this new function as:\begin{equation}
r_{k}(n)=r_{0}(n-l)+(-1)^{k}\frac{1}{\pi}\left[\sum_{j=1}^{n-l}(-1)^{n-j}\frac{s(k-1,2(n-j-k)+1)}{2j-1}+\sum_{j=1}^{l}(-1)^{n+j}\frac{s(k-2j,2(n+j-k)-1)}{2j-1}\right]\label{eq:23}\end{equation}
where $l=\lfloor\frac{k}{2}\rfloor$. We can verify that these two
solutions for $r_{k}(n)$ satisfy eq. \ref{eq:18} by direct substitution
and the use of the following identities for the function $s(n,m)$.\begin{equation}
s(n,m)=s(n-1,m)+s(n-1,m+1)+s(n,m-1)\label{eq:24}\end{equation}
\begin{equation}
4s(n,m)=s(n-1,m)+s(n+1,m)-s(n-1,m+2)-s(n+1,m-2)\label{eq:25}\end{equation}
Eq. \ref{eq:24} can be derived from the definition of $s(n,m)$ given
in eq. \ref{eq:21}. Eq. \ref{eq:25} is gotten by applying eq. \ref{eq:24}
four times.

Adding the two solutions for $r_{k}(n)$ and using $r_{k}(n)=g(n,m)$,
with $m=n-k$, gives us the following equation for the general matrix
element of the Green function:\begin{eqnarray}
g(n,m) & = & h(l)+\frac{(-1)^{m}}{4}s(n-m-1,2m)\nonumber \\
 & + & \frac{(-1)^{m}}{\pi}\sum_{j=1}^{l}(-1)^{j}\frac{s(n-m-1,2(m-j)+1)}{2j-1}\nonumber \\
 & + & \frac{(-1)^{m}}{\pi}\sum_{j=1}^{\lfloor\frac{n-m}{2}\rfloor}(-1)^{j}\frac{s(n-m-2j,2(m+j)-1)}{2j-1}\label{eq:26}\end{eqnarray}
where $l=\lfloor\frac{n+m+1}{2}\rfloor$ and $h(l)=\frac{1}{\pi}\sum_{j=1}^{l}\frac{1}{2j-1}$.

\section{conclusion}

The symmetry of the square lattice means that we have $g(n,m)=g(m,n)$.
This allowed the difference equation for $g(n,m)$ to be reduced to
a series of first order equations for the subdiagonal matrix elements.
Solving these equations gave us an expression for the general Green
function matrix element $g(n,m)$. The value of the diagonal matrix
elements $g(n,n)$ determine a unique solution. This same procedure
can also be used to find the lattice Green function for the discrete
Helmholtz equation as long as the diagonal matrix elements can be
calculated directly. An initial investigation indicates that it may
also be possible to apply this same type of procedure to cubic and
higher dimensional lattices. For a cubic lattice, the values along
the main diagonal $g(n,n,n)$ will give a unique solution to the problem,
and so on for higher dimensional lattices. Work on these problems
is currently being pursued.

\begin{acknowledgments}
The authors acknowledge the generous support of Exstrom Laboratories
and its president Istvan Hollos.\bibliographystyle{apsrev}
\bibliography{/doc/articles/lgfint/gf}

\begin{thebibliography}{4}
\expandafter\ifx\csname natexlab\endcsname\relax\def\natexlab#1{#1}\fi
\expandafter\ifx\csname bibnamefont\endcsname\relax
  \def\bibnamefont#1{#1}\fi
\expandafter\ifx\csname bibfnamefont\endcsname\relax
  \def\bibfnamefont#1{#1}\fi
\expandafter\ifx\csname citenamefont\endcsname\relax
  \def\citenamefont#1{#1}\fi
\expandafter\ifx\csname url\endcsname\relax
  \def\url#1{\texttt{#1}}\fi
\expandafter\ifx\csname urlprefix\endcsname\relax\def\urlprefix{URL }\fi
\providecommand{\bibinfo}[2]{#2}
\providecommand{\eprint}[2][]{\url{#2}}

\bibitem[{\citenamefont{Cserti}(2000)}]{cserti00}
\bibinfo{author}{\bibfnamefont{J.}~\bibnamefont{Cserti}},
  \bibinfo{journal}{Amer. J. Phys.} \textbf{\bibinfo{volume}{68}},
  \bibinfo{pages}{896} (\bibinfo{year}{2000}), \eprint{cond-mat/9909120}.

\bibitem[{exs()}]{exstrom2005}
\eprint{http://www.exstrom.com/lgf/lgfpe/lgfpe.html}.

\bibitem[{\citenamefont{Hollos}(2005)}]{hollos05_2}
\bibinfo{author}{\bibfnamefont{S.}~\bibnamefont{Hollos}},
  \emph{\bibinfo{title}{A lattice green function introduction}}
  (\bibinfo{year}{2005}), \eprint{http://www.exstrom.com/lgf/lgf.html}.

\bibitem[{\citenamefont{Hollos and Hollos}(2005)}]{hollos05_1}
\bibinfo{author}{\bibfnamefont{S.}~\bibnamefont{Hollos}} \bibnamefont{and}
  \bibinfo{author}{\bibfnamefont{R.}~\bibnamefont{Hollos}},
  \emph{\bibinfo{title}{Some square lattice green function formulas}}
  (\bibinfo{year}{2005}), \eprint{cond-mat/0508779}.

\end{thebibliography}

\end{acknowledgments}

\end{document}